# Defining Urban Boundaries by Characteristic Scales

Yanguang Chen[1], Jiejing Wang[2], Yuqing Long[1], Xiaohu Zhang[3], Xiaoping Liu[4], Xiaosong Li[5]

(1.Department of Geography, College of Urban and Environmental Sciences, Peking University, Beijing, China; 2. Department of Urban Planning and Management, School of Public Administration and Policy, Renmin University of China, Beijing, China; 3. Future Urban Mobility IRG, Singapore-MIT Alliance for Research and Technology, Singapore; 4. School of Geography and Planning, Sun Yat-sen University, Guangzhou, Guangdong, China; 5. Institute of Remote Sensing and Digital Earth, Chinese Academy of Sciences, Beijing, China.)

**Abstract**: Defining an objective boundary for a city is a difficult problem, which remains to be solved by an effective method. Recent years, new methods for identifying urban boundary have been developed by means of spatial search techniques (e.g. CCA). However, the new algorithms are involved with another problem, that is, how to determine the characteristic radius of spatial search. This paper proposes new approaches to looking for the most advisable spatial searching radius for determining urban boundary. We found that the relationships between the spatial searching radius and the corresponding number of clusters take on an exponential function. In the exponential model, the scale parameter just represents the characteristic length that we can use to define the most objective urban boundary objectively. Two sets of China's cities are employed to test this method, and the results lend support to the judgment that the characteristic parameter can well serve for the spatial searching radius. The research may be revealing for making urban spatial analysis in methodology and implementing identification of urban boundaries in practice.

## 1 Introduction

One of basic measures of a city is its size, which can be evaluated by urban population. Population is one of the central variables in the studies on urban evolution (Dendrinos, 1996), and it represents the first dynamics of city development (Arbesman, 2012). In order to determine the urban population in the proper way, we must determine the urban area. The precondition of determining the urban area effectively is to determine the urban boundary objectively. However, for a long time, the definition of urban boundary is a hard technical problem. The concept of "urban" differs from country to country (Zhou, 1995). If the areal units of cities are improper, the corresponding city population cannot be employed to measure city sizes effectively and test urban regularities such as the rank-size rule, allometric growth law, and gravity law. The levels of urbanization in different regions are also not comparable with each other. In urban geography, there exist three key concepts of cities: city proper (CP), urbanized area (UA), and metropolitan areas (MA) (Davis, 1978; Zhou, 1995). Generally speaking, the definition of "city" refers to the urbanized area. A new trend seems to be that urbanized area will be replaced by urban agglomeration, but the method of identifying the boundary of an urban agglomeration is a problem.

Geographical phenomena fall into two groups: one is with characteristic scale (scaleful group), and the other, without characteristic scale (scale-free group). The former can be described with characteristic length such as average value, standard deviation, and eigenvalue, while the latter should be described with scaling exponent such as fractal dimension (Chen, 2021). Generally speaking, it is easy to find characteristic scales for simple systems, but it is often difficult to find characteristic scales for complex systems. On the other, a complex system many follow scaling law, but it sometimes bear simple sides with some type of characteristic scales. Geographical systems comprise both scaleful and scale-free processes and patterns, which are woven into each other. If we can find the simple side and work out the corresponding characteristic scales for spatial measurement, maybe we find an effective approach to defining urban boundaries. The boundary curve of a city is termed *urban envelopes* (Batty & Longley, 1994; Longley *et al*., 1991). Based on remote sensing images, at least three approaches have been developed to determining urban envelopes for cities. The first is the city clustering algorithm (CCA) proposed by Rozenfeld *et al*. (2008; 2011), the second is the method of clustering street nodes/blocks advanced by Jiang and Jia (2011), and the third is the fractal-based method presented by Tannier *et al*. (2011). Among these methods, CCA is based on raster data, while street network clustering is based on vector data. The two methods are involved with the technology of automated spatial search based on digital maps. The key lies in how to determine the characteristic searching radius.

An interesting finding is that the relationship between the searching radius of CCA and the number of clusters of an urban agglomeration follows a negative exponential law. A characteristic length can be found from the exponential distribution. Based on the finding, this paper is devoted to exploring the methods of definition of urban boundary using the characteristic length. It should be clear that this study is based on two geographical laws: one is the distance decay law of urban density (e.g., Batty & Longley, 1994; Bussiere & Snickars, 1970; Chen, 2009; Clark, 1951; MacKinnon, 1970; Smeed, 1963; Tobler, 1970), which guarantees the feasibility of spatial search technique; and the other is the allometric scaling law of urban size and shape (e.g., Batty, 2008; Batty et al., 2008; Batty & Longley, 1994; Bettencourt, 2013; Bettencourt *et al*., 2007; Chen, 2010, 2014; Chen *et al*., 2019; Lee, 1989; Lo, 2002; Lo & Welch, 1977), which suggests that urbanized area can be employed to represent city population. By means of empirical analysis based on observational data, we will show how to calculate characteristic searching radius. The other parts of this article are organized as follows. In Section 2, the principle of spatial search for urban boundary is illuminated, and the mathematical models for characteristic scale analysis are presented. In Section 3, two sets of Chinese cities are employed to make case analyses to show how to find the characteristic radius for spatial search. In Section 4, several related questions are discussed. Finally, the discussion is concluded by summarizing the main points of this work.

## 2 Models and methods

### 2.1 Mathematical models

The CCA and its variants can be employed to determine an urban boundary line, and thus a set of spatial measurements such as area, perimeter, and cluster number can be counted. In this paper, a cluster implies the group of urban elements which are linked with one another in a digital map or remote sensing image. Changing the searching radius yield different values of spatial measurements, including urban area, street length, and number of street nodes. Many measurements increase with the increase of searching radius, but the numbers of clusters are special and decrease over searching radius. In the process of CCA, the larger the searching radius, the larger the urban area, but the more urban elements are merged at the same time. As a result, the number of urban clusters decreases. Using the datasets from spatial searching, we can find a functional relationship between the searching radius and cluster numbers. If the relationship follows a power law, we will be unable to find a characteristic length directly for the searching radius, and thus cannot determine an objective urban boundary through a simple way. If so, we cannot find an objective urban boundary in theory. On the contrary, if the relationship satisfies a function with a characteristic scale, we will be able to find a characteristic length for the searching radius, and thus determine an objective urban envelope.

This type of functions can be treated as scale functions. The scale functions include exponential function, normal function, logarithmic function, logistic function, and so on. A scale-free function such as power function follows scaling relation, while a scale function is not.

By trial and error based on observational data, we find that the number of urban clusters is most likely an exponential function of the searching radius. This is a scale function with characteristic length. According to the research of Chen *et al* (2019), in many cases, there is an exponential relationship between searching radii and corresponding measurements. So, the hypothesis to be testified in this work is that the number of clusters in a city is an exponential function of searching radius. The basic model is an exponential relation as below

$$N(s) = N_0 e^{-s/s_0}, \quad (1)$$

where $s$ denotes the length of searching radius, $N(s)$ refers to the number of clusters, $N_0$ is a proportionality constant, and $s_0$ is the scale parameter indicating some characteristic length. Another possible function is the normal function, that is

$$N(s) = N_0 e^{-s^2/(2s_0^2)}, \quad (2)$$

which is in fact the variant of Gaussian function and can be treated as a quadratic exponential function. The symbols are the same as in equation (1). In practice, the exponential function and the normal function can be unified into a general form as follows

$$N(s) = N_0 e^{-s^\sigma/(\sigma s_0^\sigma)}, \quad (3)$$

where $\sigma$ denotes a latent scaling exponent, which comes between 1/2 and 2 (Chen, 2010). Equation (3) can be regarded as a fractional exponential function. If $\sigma$=1, we will have an exponential function; and if $\sigma$=2, then we will have a normal function. In above equations, the scale parameter $s_0$ represents characteristic length of the searching radius. In theory, the scale parameter $s_0$ is related to the average value of the searching radius. If we find the $s_0$ value, we will find the most appropriate searching radius for defining an urban boundary. As a result, the most objective urban envelopes can be identified from digital maps by means of GIS technology.

## 2.2 The method of spatial search for urban envelope

In this work, the study object of CCA is remote sensing images of urban agglomerations in China. The general idea of city clustering is to aggregate land patches within a distance threshold $s$.

Although the data used in this study are vector files, the algorithm itself can be formulated in raster format. If the distance of two patches is smaller than $s$, the two patches will be identified as in the same cluster. The larger the distance threshold $s$, the less the formed clusters. Number of clusters will be reduced to 1 if the threshold is larger than the radius of study area. Since the clustering needs to be done in a set of different radii, iteratively joining nearby polygons could be too computationally intensive, especially in a large study area. We implemented an efficient method to identify clusters. Firstly, we calculated distances between patches. But instead of computing every pair of them, we filtered out those which distances were larger than the maximum distance threshold using spatial indexing. This generated a sparse matrix $M$ with its element $d_{ij}$ denoting the distance of patch $i$ to patch $j$. We then performed clustering over a series of distance thresholds. For each distance threshold $s$, we constructed a network which nodes were patches and two nodes (patches) were connected if their distance is smaller than $s$. The network would have many isolated sub-networks and nodes (patches) in each sub-network were identified as in the same cluster.

# 3 Empirical results

## 3.1 Material and methods

The datasets are the land-use patches interpreted from Landsat satellite imagery of China. The images were classified into X categories, from which the built-up areas are extracted for our study. First of all, 24 Chinese cities are employed to show how to find a characteristic length of searching radius to define urban boundary. These 24 cities are scattered on all over China (Figure 1). The typical regions consist of National capital (i.e., Beijing), Yangtze River Delta (e.g., Hangzhou, Nanjing, Shanghai, Wuxi), Pearl River Delta (e.g. Guangzhou), Central Plain (e.g., Kaifeng, Luoyang), Central south China (e.g., Changsha), Shandong Peninsula (Rizhao), northeast China (e.g. Changchun, Fuxin, Liaoyuan), northwest China (e.g., Changzhi, Yinchuan), southwest Chian (e.g., Chengdu, Kunming). Where urban hierarchy is concerned, these cities comprise megacities (megalopolis), large cities (metropolis), medium-sized cities, and small cities (Table 1). The selection of cities is similar to zoning and stratified sampling. In terms of space, this sampling should ensure that different places in China are involved. In terms of hierarchy, the sampling should ensure the different size levels of cities are represented. A Zipf's distribution of cities proved to be equivalent to a self-similar hierarchy of cities with cascade cities (Wang and Chen, 2021). The 24 cities were choose from different size classes of urban hierarchy by taking places in space into account. The original remote sensing images are of vector format and belong to 2000 and 2010, respectively. The spatial datasets were extracted and processed by ArcGIS from the digital maps of urban land use (see attached Files S1 and S2).

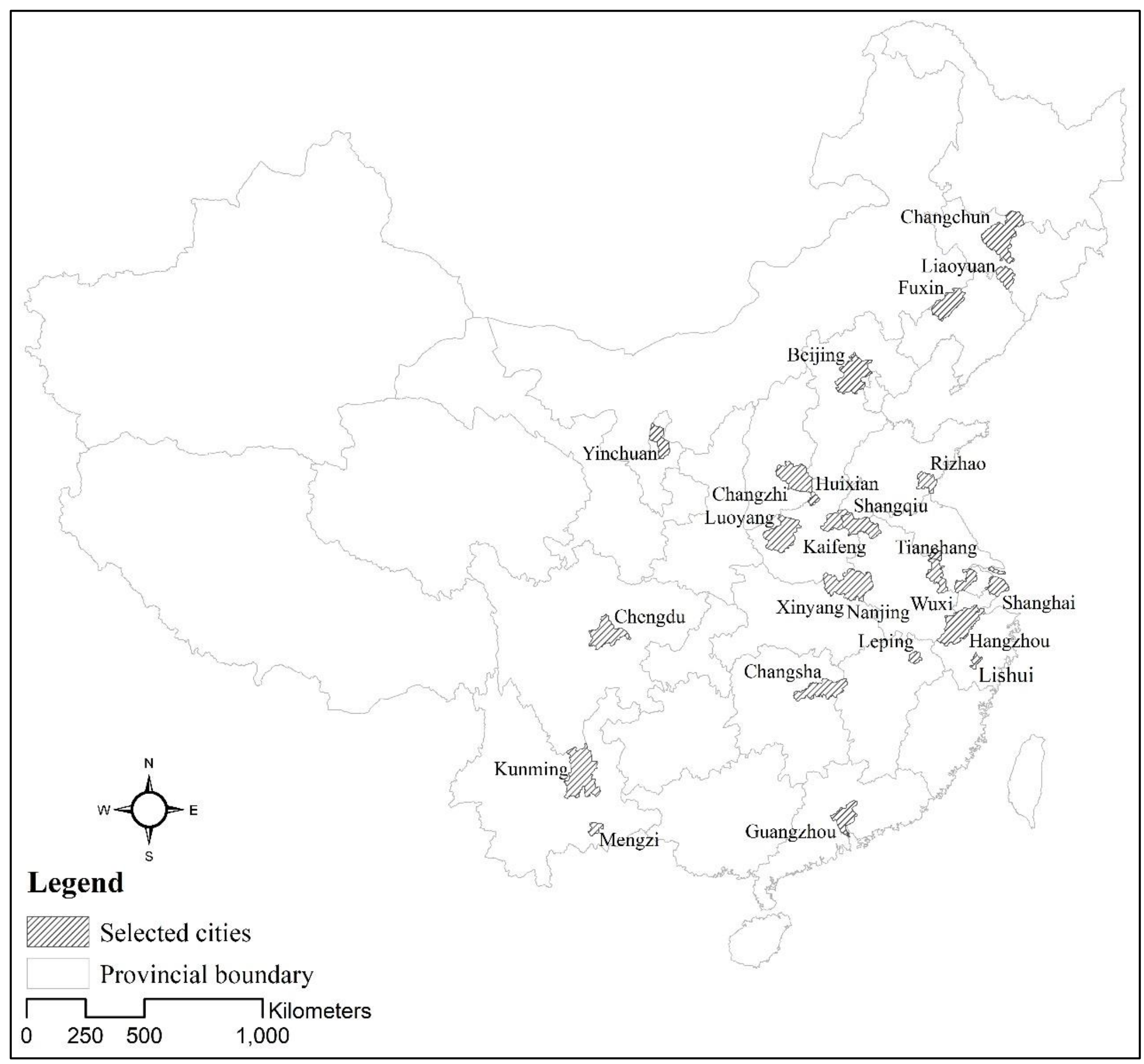

**Figure 1 A spatial sample of 24 Chinese cities which are taken as examples of characteristic radius analysis**

A set of searching radii can be given to generate datasets of boundary curves of a city. Based on an urban boundary, three measurements can be obtained, that is, number of clusters, the area of the largest cluster, and the total area of all the clusters. We can write a computer program of ArcGIS to perform the search and design a cycle to control the searching radius. By repeated tests, the minimum searching radius is taken $s$=50 meters, and the step length of radius change is set as $\Delta s$=10 meters. The maximum searching radius ($s_{max}$) depends on urban shape, network pattern, map layout, and so on. For each selected city, we use the method descripted in the above section to perform the spatial searching. Where the national capital of China is concerned, the larger searching radii leads to larger urban clusters (Appendix 1). The cases of other cities are similar to Beijing (Figure 2). The cluster number, cluster area, and the area of the largest cluster can be automatically extracted by the computer program during each searching cycle. Thus we will have three datasets for a series of variable urban boundaries of each city.

**Table 1 The 24 cities as a sample from the hierarchy of Chinese cities**

| Size level | Cities (6 ones in each level) | Size |
|---|---|---|
| **Megacity** | Beijing, Chengdu, Guangzhou, Hangzhou, Nanjing, Shanghai | >5000 |
| **Large city** | Changchun, Changsha, Kunming, Luoyang, Wuxi, Yinchuan | 1000-5000 |
| **Medium city** | Changzhi, Fuxin, Kaifeng, Rizhao, Shangqiu, Xinyang | 500-1000 |
| **Small city** | Huixian, Leping, Liaoyuan, Lishui, Mengzi, Tianchang | <500 |

**Note**: The unit of city size is thousand people.

The linear regression analysis based the least squares method can be employed to estimate the values of parameters. Taking natural logarithms on both sides of equation (3) yields

$$\ln N(s) = \ln N_0 - \frac{s^\sigma}{\sigma s_0^\sigma} = a - bs^\sigma , \tag{4}$$

in which the parameters are $a=\ln N_0$ and $b=1/(\sigma s_0^\sigma)$, respectively. Thus, the characteristic radius can be given by

$$s_0 = (\sigma b)^{-1/\sigma} . \tag{5}$$

If $\sigma=1$ as given, then equation (4) will return to an ordinary linear relation such as

$$\ln N(s) = \ln N_0 - \frac{s}{s_0} = a - bs , \tag{6}$$

Thus we have $s_0=1/b$. If the relation between the searching radius and cluster number follow the exponential decay law, we can use the linear regression analysis based on equation (6) to evaluate the characteristic radius $s_0$. If the relation follow the fractional exponential law, we can find the value of the latent scaling exponent $\sigma$ by means of the cut-and-try method. As soon as the $\sigma$ value is determined, we can use equation (4) to make a linear regression analysis and find the characteristic radius $s_0$.

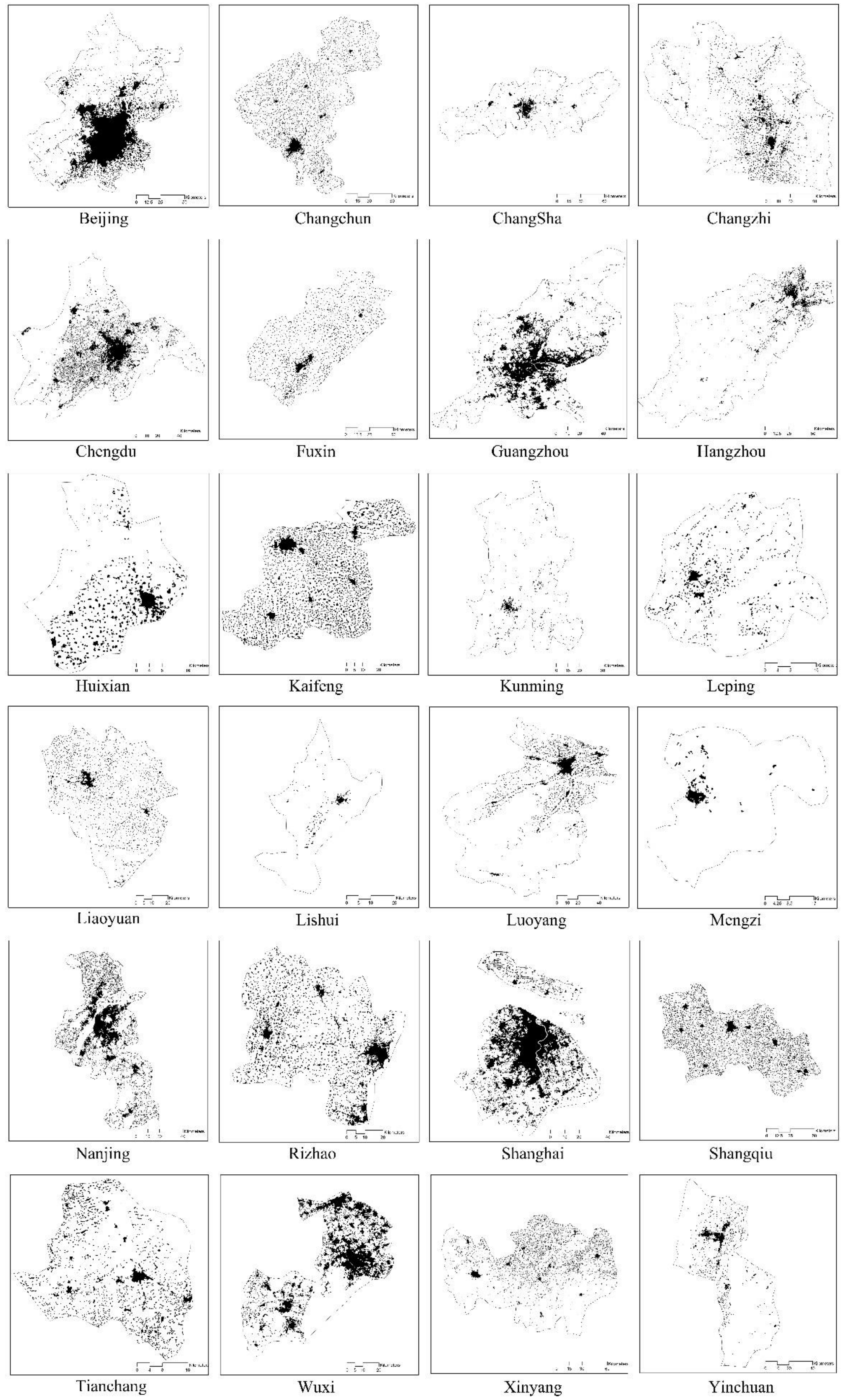

**Figure 2 The sketch maps of the spatially identified results of the 24 Chinese cities**

[**Note**: The area of a city depends on the searching radius. Longer searching radius results in larger urban area.]

### 3.2 Results

The exponential function and the fractional exponential function can applied to the spatial searching processes of Chinese cities. Concretely speaking, equations (4) and (6) can be fitted to the datasets of the 24 Chinese cities. For example, for Beijing in 2010, the relation between searching radii and cluster numbers can be well described by the common exponential function. By the least squares computation, the model can be built as follows

$$\hat{N}(s) = 2416.8043e^{-0.0017s}, \tag{7}$$

where the hat "^" indicates that the result is a predicted value rather than an observed value. The coefficient of determination is about $R^2$=0.9992. The decay coefficient is $b$≈0.00173992, thus the characteristic radius of spatial search is estimated as $s_0=1/b≈574.7391$ (Figure 3). In contrast, fitting power function to the same dataset, the goodness of fit is about $R^2$=0.8863. The scattered points cannot be well matched with the trend line based on power law relation. Among various possible scale functions, the negative exponential function is a very good model for the relationship between searching radius and cluster number. Where the spatial searching is concerned, all the 24 cities can be approximately modeled by the negative exponential function (See Files S1 and S2). Thus the characteristic searching radius can evaluated by the least squares regression based on equation (1). According to the exponential model, for the 24 cities in 2000, the average characteristic searching radius is about $\hat{s}_{0(2000)}$=479.4751; for these cities in 2010, the average value is about $\hat{s}_{0(2010)}$=463.0777 (Table 2). Here $\hat{s}_0$ refers to the mean of the characteristic searching radii of different cities.

**Table 2 The regression coefficients, characteristic searching radii and the goodness of fit of the 24 Chinese cities based on exponential function**

| City | 2000 | | | 2010 | | |
|---|---|---|---|---|---|---|
| | $b$ | $s_0$ | $R^2$ | $b$ | $s_0$ | $R^2$ |
| **Beijing** | 0.002058 | 485.8544 | 0.9944 | 0.001740 | 574.7391 | 0.9992 |
| **Changchun** | 0.003046 | 328.3512 | 0.9576 | 0.002994 | 333.9824 | 0.9511 |
| **Changsha** | 0.001528 | 654.5316 | 0.9819 | 0.001522 | 656.9094 | 0.9805 |
| **Changzhi** | 0.001195 | 836.5261 | 0.9822 | 0.001694 | 590.2909 | 0.9990 |
| **Chengdu** | 0.003978 | 251.3687 | 0.9471 | 0.003887 | 257.2394 | 0.9487 |
| **Fuxin** | 0.001527 | 654.8017 | 0.9879 | 0.001480 | 675.8264 | 0.9773 |
| **Guangzhou** | 0.001895 | 527.6015 | 0.9941 | 0.001684 | 593.8842 | 0.9951 |

| **Hangzhou** | 0.001779 | 562.1262 | 0.9799 | 0.001746 | 572.7869 | 0.9810 |
|---|---|---|---|---|---|---|
| **Huixian** | 0.002493 | 401.1521 | 0.9579 | 0.002462 | 406.1788 | 0.9797 |
| **Kaifeng** | 0.004095 | 244.1824 | 0.9498 | 0.004308 | 232.1424 | 0.9582 |
| **Kunming** | 0.001070 | 934.3961 | 0.9870 | 0.001001 | 998.9711 | 0.9899 |
| **Leping** | 0.001868 | 535.2374 | 0.9978 | 0.001910 | 523.6562 | 0.9974 |
| **Liaoyuan** | 0.002257 | 443.0955 | 0.9714 | 0.003078 | 324.8578 | 0.9634 |
| **Lishui** | 0.001375 | 727.5108 | 0.9865 | 0.001552 | 644.4254 | 0.9821 |
| **Luoyang** | 0.002181 | 458.4779 | 0.9956 | 0.002598 | 384.8996 | 0.9928 |
| **Mengzi** | 0.000825 | 1211.5485 | 0.9734 | 0.000866 | 1155.3882 | 0.9503 |
| **Nanjing** | 0.005397 | 185.2854 | 0.9972 | 0.005248 | 190.5499 | 0.9973 |
| **Rizhao** | 0.002173 | 460.1001 | 0.9624 | 0.002654 | 376.7287 | 0.9776 |
| **Shanghai** | 0.003404 | 293.7530 | 0.9959 | 0.003276 | 305.2242 | 0.9976 |
| **Shangqiu** | 0.008037 | 124.4241 | 0.9845 | 0.007993 | 125.1165 | 0.9747 |
| **Tianchang** | 0.003609 | 277.1104 | 0.9979 | 0.003709 | 269.6406 | 0.9976 |
| **Wuxi** | 0.004816 | 207.6537 | 0.9975 | 0.004290 | 233.0953 | 0.9954 |
| **Xinyang** | 0.004204 | 237.8845 | 0.9871 | 0.004117 | 242.8776 | 0.9846 |
| **Yinchuan** | 0.002153 | 464.4294 | 0.9971 | 0.002250 | 444.4543 | 0.9976 |
| **Average** | **0.002790** | **479.4751** | **0.9818** | **0.002836** | **463.0777** | **0.9820** |

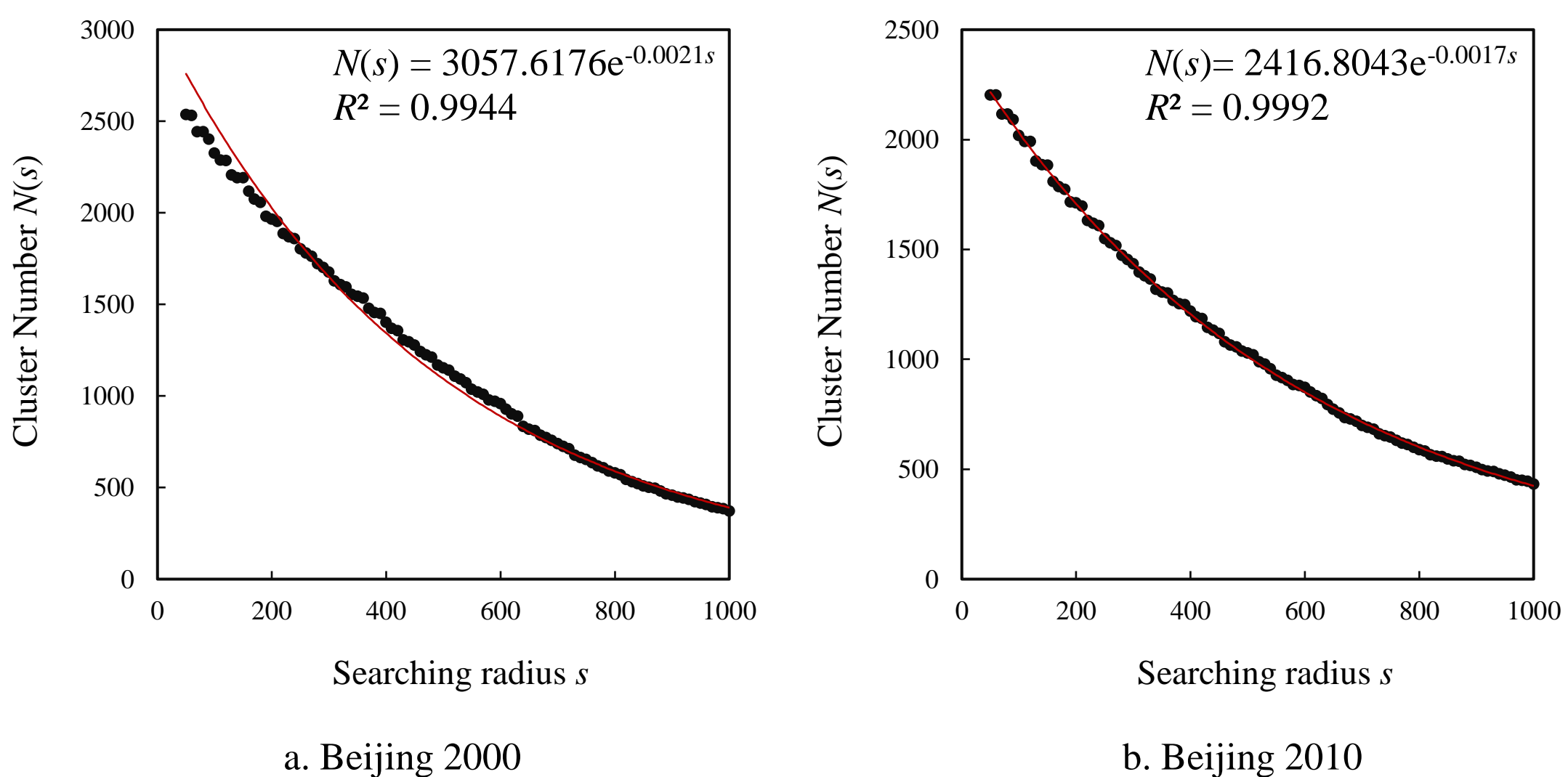

a. Beijing 2000 b. Beijing 2010

**Figure 3 The exponential decay relations between the searching radii and the numbers of clusters of Beijing city**

(**Note**: For Beijing in 2000, the negative exponential function is not the best model, the most probable model is a fractional exponential model with a latent scaling exponent $\sigma$=1.25. However, in 2010, the common negative exponential function is the most advisable model for Beijing city. Power function cannot be well fitted to the two datasets. The goodness of fit for power function is $R^2$=0.8423 for 2000 dataset and 8863 for 2010 dataset, respectively. The results are seriously biased. )

However, not all cities can be well described with the common exponential decay function at all times. For example, the cities of Changchun and Kaifeng satisfy the normal function. Great majority of cities follow the fractional exponential decay law. Where Beijing in 2000 is concerned, it should be described with the generalized exponential function such as

$$\hat{N}(s) = 2610.4595e^{-0.0004s^{1.25}}, \tag{8}$$

where the latent scaling exponent is about $\sigma$=1.25. The goodness of fit is about $R^2$=0.9992. The decay coefficient is $b$≈0.0003507, thus the characteristic radius of spatial search is estimated as $s_0$ = $(1.25*0.0003507)^{(-1/1.25)} \approx 485.8737$. Luoyang city in 2010 is similar to Beijing city in 2000 (Figure 4). According to the fractional exponential model, for the 24 cities in 2000, the average characteristic searching radius is about $\hat{s}_{0(2000)}$=528.7216; for them in 2010, the average value of typical searching radii is about $\hat{s}_{0(2010)}$=569.6380 (Table 3).

**Table 3 The regression coefficients, characteristic searching radii and the goodness of fit of the 24 Chinese cities based on fractional exponential function**

| City | 2000 | | | | 2010 | | | |
|---|---|---|---|---|---|---|---|---|
| | $\sigma$ | $b$ | $s_0$ | $R^2$ | $\sigma$ | $b$ | $s_0$ | $R^2$ |
| **Beijing** | 1.250 | 0.000351 | 485.8737 | 0.9992 | 1.000 | 0.001740 | 574.7391 | 0.9992 |
| **Changchun** | 1.900 | 0.000006 | 408.6553 | 0.9990 | 2.000 | 0.000003 | 419.5907 | 0.9991 |
| **Changsha** | 0.600 | 0.029259 | 843.3116 | 0.9982 | 0.600 | 0.029174 | 847.3903 | 0.9983 |
| **Changzhi** | 1.500 | 0.000036 | 704.3982 | 0.9995 | 1.000 | 0.001694 | 590.2909 | 0.9990 |
| **Chengdu** | 0.333 | 0.680013 | 85.8640 | 0.9960 | 0.350 | 0.573566 | 98.2707 | 0.9969 |
| **Fuxin** | 1.400 | 0.000091 | 604.6122 | 0.9977 | 1.600 | 0.000022 | 606.1060 | 0.9989 |
| **Guangzhou** | 1.000 | 0.001895 | 527.6015 | 0.9941 | 1.000 | 0.001684 | 593.8842 | 0.9951 |
| **Hangzhou** | 0.600 | 0.034105 | 653.2190 | 0.9983 | 0.600 | 0.033453 | 674.5782 | 0.9984 |
| **Huixian** | 1.850 | 0.000007 | 451.2435 | 0.9966 | 1.500 | 0.000073 | 435.0433 | 0.9978 |
| **Kaifeng** | 2.000 | 0.000004 | 358.5174 | 0.9994 | 1.800 | 0.000016 | 330.5710 | 0.9985 |
| **Kunming** | 0.700 | 0.009580 | 1273.5779 | 0.9957 | 0.750 | 0.006170 | 1296.7000 | 0.9958 |
| **Leping** | 1.000 | 0.001868 | 535.2374 | 0.9978 | 1.000 | 0.001910 | 523.6562 | 0.9974 |
| **Liaoyuan** | 1.700 | 0.000017 | 469.7532 | 0.9997 | 1.800 | 0.000012 | 398.8178 | 0.9993 |
| **Lishui** | 1.250 | 0.000234 | 671.0777 | 0.9914 | 1.400 | 0.000093 | 597.3745 | 0.9936 |
| **Luoyang** | 1.000 | 0.002181 | 458.4779 | 0.9956 | 1.250 | 0.000443 | 403.1791 | 0.9983 |
| **Mengzi** | 0.700 | 0.007392 | 1844.5408 | 0.9828 | 0.500 | 0.036380 | 3022.2644 | 0.9773 |
| **Nanjing** | 1.000 | 0.005397 | 185.2854 | 0.9972 | 1.000 | 0.005248 | 190.5499 | 0.9973 |
| **Rizhao** | 1.800 | 0.000008 | 483.8662 | 0.9989 | 1.600 | 0.000040 | 420.6628 | 0.9992 |
| **Shanghai** | 1.000 | 0.003404 | 293.7530 | 0.9959 | 1.000 | 0.003276 | 305.2242 | 0.9976 |
| **Shangqiu** | 1.250 | 0.001370 | 163.3755 | 0.9896 | 1.500 | 0.000238 | 198.5161 | 0.9911 |
| **Tianchang** | 1.000 | 0.003609 | 277.1104 | 0.9979 | 1.000 | 0.003709 | 269.6406 | 0.9976 |

| **Wuxi** | 1.000 | 0.004816 | 207.6537 | 0.9975 | 1.000 | 0.004290 | 233.0953 | 0.9954 |
|---|---|---|---|---|---|---|---|---|
| **Xinyang** | 1.000 | 0.004204 | 237.8845 | 0.9871 | 0.750 | 0.025384 | 196.7128 | 0.9909 |
| **Yinchuan** | 1.000 | 0.002153 | 464.4294 | 0.9971 | 1.000 | 0.002250 | 444.4543 | 0.9976 |
| **Average** | **1.160** | **0.033000** | **528.7216** | **0.9959** | **1.125** | **0.030453** | **569.6380** | **0.9962** |

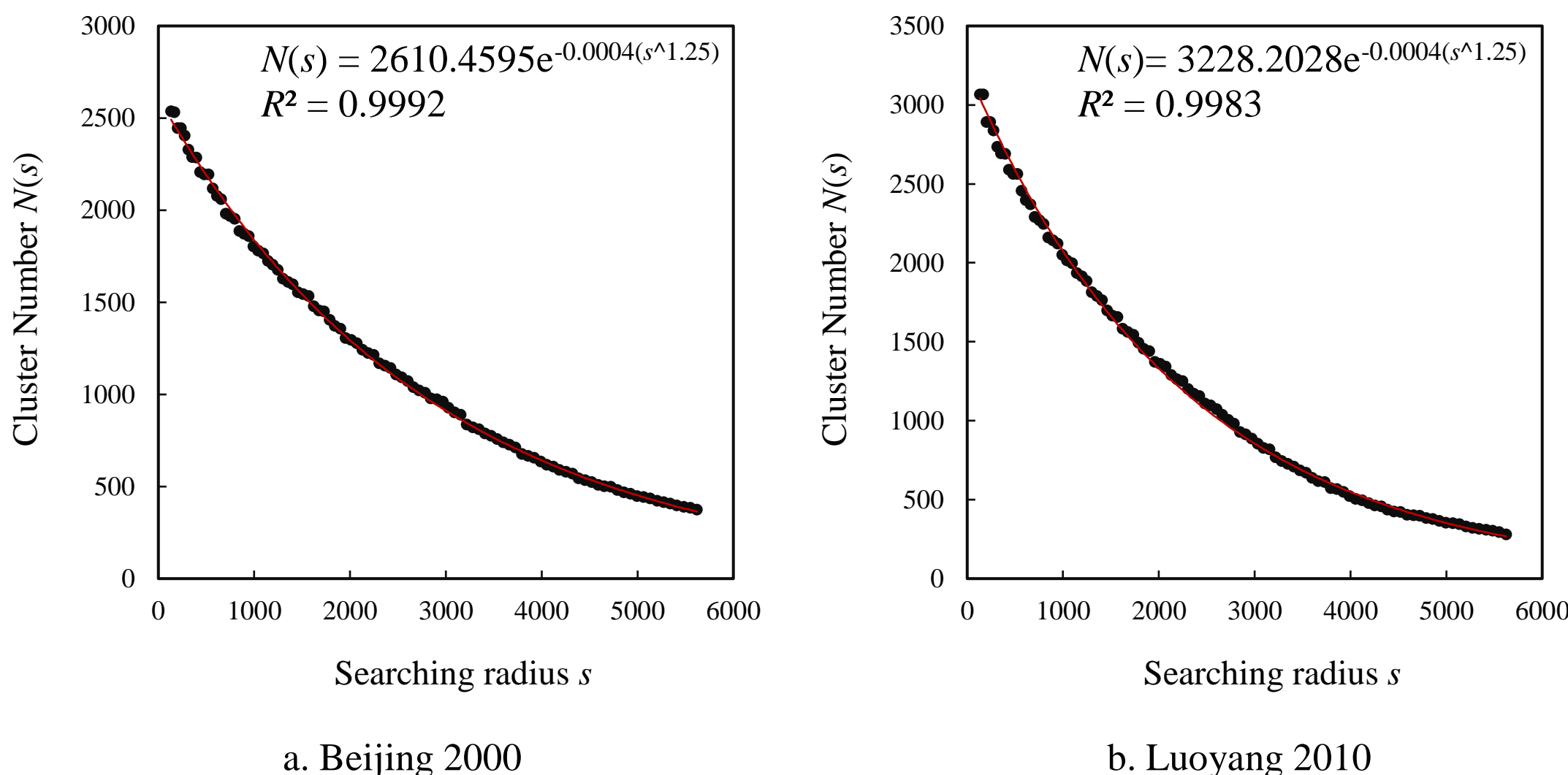

a. Beijing 2000 b. Luoyang 2010

**Figure 4 The fractional exponential decay relations between the searching radii and the numbers of clusters of Beijing city ($\sigma$=1.25)**

(**Note**: For both Beijing in 2000 and Luoyang city in 2010, the most probable latent scaling exponent is $\sigma$=1.25.)

There are two ways of defining boundaries for cities. One is for an individual city, and the other is for all cities in a region (country). If we want to find the most advisable boundary for a particular city, we can use the fractional exponential function to find the characteristic searching radius of the city at given time (e.g., Table 2). However, for different cities at different times, the values of the characteristic searching radius are always different. Thus it is difficult to define strictly comparable urban boundaries for a set of cities. In this instance, we can calculate the average value of characteristic searching radii of a sample of cities. Unfortunately, another problem arises: the average value is not often stable. A solution to this problem is to remove the outliers in a sample using the double standard derivation before computing the average value. Based on the significance level $\alpha$=0.05, an acceptable characteristic searching radius $s_0$ should fall between average value plus double standard deviation and average value minus double standard deviation, that is

$$\hat{s}_0 - 2sd \le s_0 \le \hat{s}_0 + 2sd\,,$$

where $\hat{s}_0$ denotes the average value (mean) of the characteristic searching radius, and $sd$ refers to the standard deviation. Otherwise, we have 95% of confidence level to treat the $s_0$ value as an outlier. After removing the outliers step by step, we can find an acceptable $\hat{s}_0$ value.

The effective average value of the characteristic searching radii can be approached step by step. Taking the dataset of the 24 cities in 2000 based on exponential model as an example, the procedure of finding a stable average value is as follows. Step 1, calculate the average value $\hat{s}_0$ and the standard deviation $sd$. If $s_0^{(k)}< \hat{s}_0-2sd$ or $s_0^{(k)}> \hat{s}_0+2sd$, remove it ($k$=1, 2,…,24). In this step, the characteristic searching radius of Mengzi city is an outlier and should be removed (Figure 5(a), Table 4). Step 2, recalculate the average value $\hat{s}_0$ and the standard deviation $sd$. If $s_0^{(k)}< \hat{s}_0-2sd$ or $s_0^{(k)}> \hat{s}_0+2sd$, remove it ($k$=1, 2,…, 23). In this step, the characteristic radius of Kunming city is an outlier and should be deleted (Figure 5(b), Table 4). Step 3, recalculate the average value $\hat{s}_0$ and the standard deviation $sd$ once again. If $s_0^{(k)}< \hat{s}_0-2sd$ or $s_0^{(k)}> \hat{s}_0+2sd$, remove it ($k$=1,2,…,22). In this step, the characteristic radius of Changzhi city is an outlier and should be eliminated (Figure 5(c), Table 4). Step 4, repeat above calculation. If $s_0^{(k)}< \hat{s}_0-2sd$ or $s_0^{(k)}> \hat{s}_0+2sd$, remove it ($k$=1,2,…,21). In this step, all the characteristic radii of the 21 remaining cities fall into the range from $\hat{s}_0-2sd$ to $\hat{s}_0+2sd$, and the final average value and standard deviation are 405.9491 and 170.3990 (Figure 5(d), Table 4).

**Table 4 An example of removing outliers by average values and double standard deviations of the 24 cities in 2000 (common exponential model)**

| Statistics\Step | Step1 | Step2 | Step3 | Step4 |
|---|---|---|---|---|
| **Average value** | 479.4751 | 447.6458 | 425.5208 | 405.9491 |
| **Standard deviation** | 260.8192 | 213.7735 | 189.9480 | 170.3990 |
| **Lower limit (mean-2*sd)** | -42.1634 | 20.0988 | 45.6247 | 65.1512 |
| **Upper limit (mean+2*sd)** | 1001.1136 | 875.1929 | 805.4169 | 746.7470 |
| **Outlier** | 1211.5485 | 934.3961 | 836.5261 | No outlier |
| **Exceptional City** | Mengzi | Kunming | Changzhi | No excepted city |

Based on the two models and two years, we have four datasets. Using the similar method, we can address all these datasets (Table 5). The other results are as below. For the dataset of 2010 based on exponential decay model, the average value and standard deviation are 407.2503 and 170.3565 (the outliers are Mengzi and Kunming); For the dataset of 2000 based on fractional exponential decay model, the average value and standard deviation are 415.6138 and 176.5345 (the outliers are Mengzi, Kunming and Changsha); For the dataset of 2010 based on fractional exponential decay model, the average value and standard deviation are 404.9980 and 166.6497 (the outliers are Mengzi, Kunming and Changsha). The most advisable characteristic searching radius can be regarded as about 410, and the corresponding standard deviation is about 170.

**Table 5 The final average values, standard deviations, and the lower and upper limits of the characteristic searching radii**

| Statistics\model | Exponential decay | | Fractional exponential decay | |
|---|---|---|---|---|
| | 2000 | 2010 | 2000 | 2010 |
| **Average value** | 405.9491 | 407.2503 | 415.6138 | 404.9980 |
| **Standard deviation** | 170.3990 | 170.3565 | 176.5345 | 166.6497 |
| **Lower limit (mean-sd)** | 235.5502 | 236.8938 | 239.0793 | 238.3482 |
| **Upper limit (mean+sd)** | 576.3481 | 577.6067 | 592.1483 | 571.6477 |

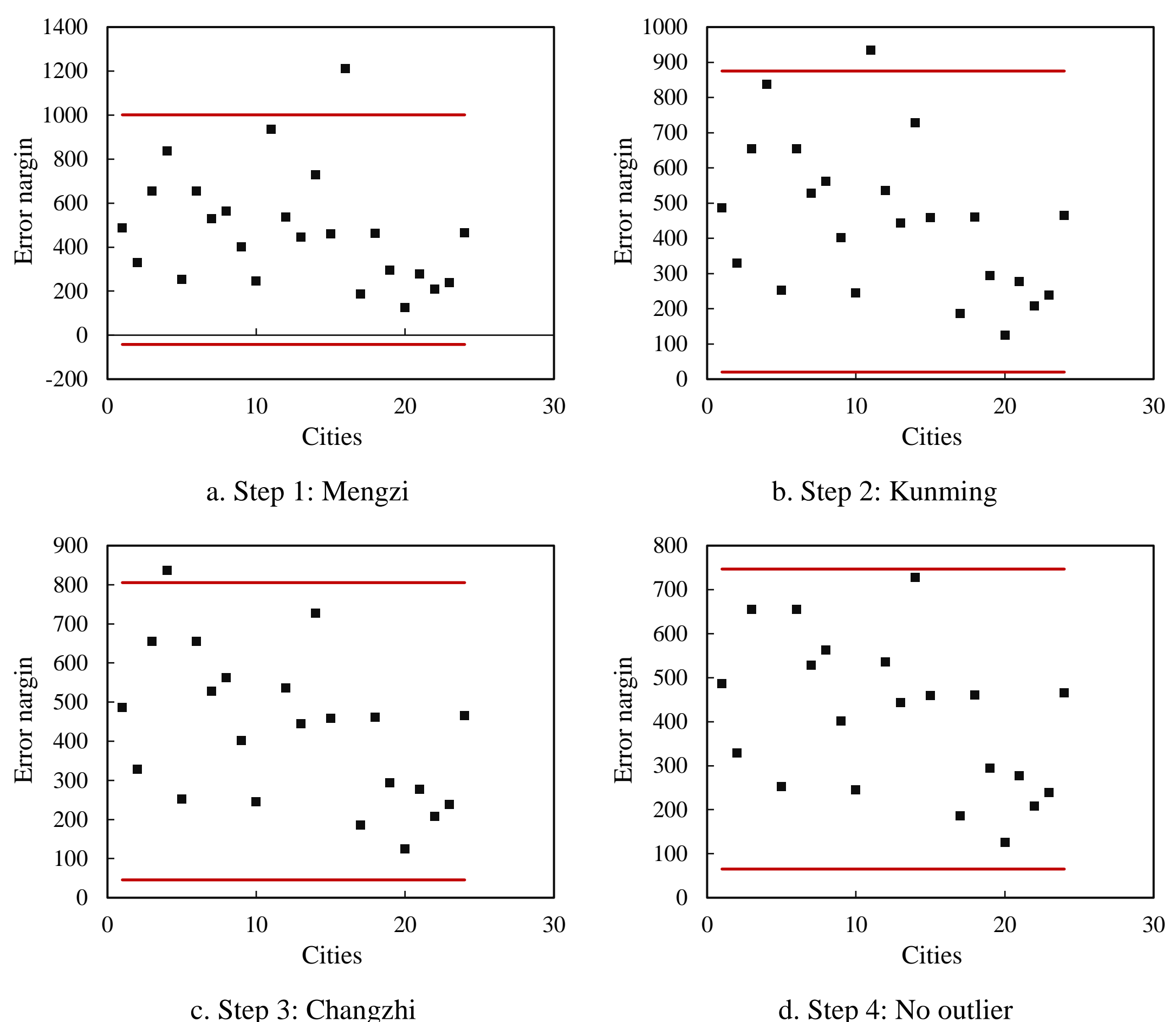

a. Step 1: Mengzi

b. Step 2: Kunming

c. Step 3: Changzhi

d. Step 4: No outlier

**Figure 5 The process and patterns of removing outliers by average values and double standard deviations of the 24 cities in 2000**

(**Note**: The model is common exponential decay function. In step 1, the city of Mengzi is an outlier; in step 2, Kunming is an outlier; in step 3, Changzhi is an outlier; in step 4, no outlier, and the process is over.)

The basic properties of geographical systems are spatial heterogeneity and hierarchical heterogeneity. The spatial heterogeneity implies spatial nonstationarity, that is, different places bear different probability structure. Similarly, the hierarchical heterogeneity indicates hierarchical nonstationarity, that is, different hierarchical levels bear different probability structure. Probability can be described by average values, standard deviation, covariance and other statistics. This suggests that different places and different size levels bear may different characteristic searching radii for defining urban boundaries. Our sample size is too small, only, 24 cities. It is not proper to examine the spatial difference of characteristic radii through this small sample. But we can preliminarily estimate the average values of different size levels in terms of Table 1. The results are shown in Table 6. Based on exponential model, the characteristic radii differences at different size levels and in different years seem to be not significant. Based on fractional exponential model, the characteristic radii values show a trend of decreasing over time and with the growth of city sizes. This suggests that the fractional exponential model is suitable for finding the characteristic searching radii for individual cities, while the exponential model is more suitable for finding the average characteristic radius approximately for a system of cities in a geographical region.

**Table 6 The average values of the characteristic searching radii at different size levels**

| **Size level** | **Based on exponential model** | | **Based on fractional exponential model** | |
|---|---|---|---|---|
| | 2000 | 2010 | 2000 | 2010 |
| **Megacity** | 384.3315 | 415.7373 | 371.9328 | 406.2077 |
| **Large city** | 422.6888 | 410.6682 | 384.8041 | 375.0799 |
| **Medium city** | 344.2785 | 373.8304 | 425.4423 | 390.4766 |
| **Small city** | 476.8212 | 433.7518 | 480.8844 | 444.9065 |
| **Mean of means** | 407.0300 | 408.4969 | 415.7659 | 404.1677 |
| **Standard deviation of means** | 56.4771 | 25.1443 | 49.0375 | 29.9853 |

**Note**: The means of four groups were calculated after removing outliers in Tables 2 and 3. The last two lines represent the means of means and the standard deviation of the means.

## 3.3 Further application to an system of cities

The above calculations and analyses are based on a sample of 24 Chinese cities. These cities have no significant relation to each other. Now, the characteristic length method can be applied to a system of cities in China, Jing-Jin-Ji region, from which we can obtain new insight in urban definition. Jing-Jin-Ji region is also termed Beijing-Tianjin-Hebei, including Beijing Municipality (ab. "Jing"), Tianjin Municipality (ab. "Jin"), and part of Hebei Province (ab. "Ji"). The Jing-Jin-Ji urban system comprises 35 cities in 2010 (Appendix 2). The original images are of raster format, differing from the previous example. Using variable searching radii such as 20, 25, and 30, we can abstract different urban agglomerations for each city in 2000, 2005, and 2010. The relationships between searching radii and cluster numbers of all these cities follow exponential decay law, and can be modeled or approximately modeled by the negative exponential function (Appendix 3). The scale parameter suggests the characteristic searching radius, $s_0$. A finding is that the characteristic searching radius becomes shorter and shorter along with urban growth. The average searching radii are 539.9886 m for 2000, 510.9694 m for 2005, and 476.9993 m for 2010, respectively. After removing the outliers from the datasets by average values and double standard deviations, the average values become 490.368 m for 2000, 471.9630 m for 2005, and 443.0722 m for 2010, respectively. Urban growth implies space filling, and thus urban density increases over time. For a city, the higher the urban density is, the shorter the characteristic searching radius will be.

Another finding is that the characteristic radius depends on the format of remote sensing image and the method of data processing. In both the sample of 24 Chinese cities and the population (universe) of 35 Jing-Jin-Ji cities, the patterns of Beijing city in 2000 and 2010 are taken into consideration. However, the results for 2010 are significantly different. The former result is 574.7391 (increase), while the latter result is 366.3731 (decrease). The first results are based on vector data (Table 2), while the second results are based on raster data (Appendix 3). For the first results, the searching radii are $s$=50, 60, 70, …, 1000, and the step length is $\Delta s$=10; For the second results, the searching radii are $s$=20, 25, 30,…, 500, and the step length is $\Delta s$=5. This suggests that only based on the same data format and the same data processing method, comparable results can be gained for the definition of urban agglomerations and urban boundaries.

## 4 Discussion

The key to define an objective urban boundary is to find a characteristic searching radius. In fact, scientific research should proceed first by describing a system and later by understanding the mechanism (Gordon, 2005; Henry, 2002). The precondition of effective description is to find the characteristic length of a thing. If an urban pattern follows power laws, it has no characteristic scale, and cannot be simply described with the traditional method; if an urban pattern satisfies an exponential distribution, it has a characteristic length, and can be effectively described in a simple way. Fortunately, a power law can be decomposed into two exponential laws, from which we may find characteristic length to define urban boundaries (Chen *et al*, 2019). As indicated above, the scale parameter of an exponential distribution model is just the characteristic length indicating the average value. In this paper, we reveal a generalized exponential relation between the spatial searching radii and the numbers of clusters consisting of land patches. In this case, the scale parameter $s_0$ give the characteristic searching radius. Based on the characteristic radius, we can define a relatively objective urban boundary. Thus, urban area and city size can be objectively determined by the objective urban envelope. For individual cities, the characteristic radii can be calculated one by one; for a system of cities, the average characteristic radius can be computed step by step using average values and double standard deviations.

The novelty of this paper is to reveal that the characteristic length of spatial searching radius can be utilized to define urban boundaries objectively. The significant shortcoming of this study lies in that only exponential decay and fractional exponential decay are taken into account. Geographical systems are different from classical physical systems. Geographical laws are not iron laws, and a geographical process or pattern can be modeled with more than one mathematical equation. Several functions can be employed to describe the relationships between the searching radii and cluster numbers, but the exponential function is the basic one and the most probable one. There are many types of spatial and probability distributions with characteristic lengths such as exponential distribution, logarithmic distribution, normal distribution, lognormal distribution, and so on. The scale-free distributions are mainly power-law distributions and latent power-law distributions. Among various distribution functions with characteristic scales, the exponential function is the simplest one. In fact, many distribution functions with characteristic scales can be transformed into an exponential function (Table 7). Normal distribution can be treated as generalized exponential distribution. The inverse function of logarithmic function is just an exponential function. It is easy to convert a sigmoid function into an exponential function.

**Table 7 Three functions and the corresponding characteristic searching radii (examples)**

| Function | Transformation | Characteristic | Parameter |
| --- | --- | --- | --- |

| | | **radius** | |
|---|---|---|---|
| $N(s)=a-b\ln(s)$ | $s=s_b e^{-N(s)/b}$ | $s_0=e^{a/b-1}$ | $a$, $b$, $s_b$=exp($a/b$) |
| $N(s)=\frac{N_{\max}}{1+ae^{bs}}$ | $\frac{N_{\max}}{N(s)}-1=ae^{bs}$ | $s_0=1/b$ | $a$, $b$, $N_{\max}$ |
| $N(s)=ae^{-bs}s^{-c}$ | $N(s)s^c=ae^{-bs}$ | $s_0=1/b$ | $a$, $b$, $c$ |
| …… | …… | …… | …… |

This paper is devoted to exploring the method of urban boundary identification based on scale functions instead of power law. If the spatial distribution follows a power law indicative of the scaling process without breaking, we will be unable to find an objective urban agglomeration or urban boundary in principle. However, in practice, we have at least three ideas to solve the problem. First, scaling decomposition. A power law can be decomposed into two exponential laws. Maybe one of exponential laws can be applied to defining urban boundaries (Chen *et al*, 2019). Second, scaling breaking. In the real world, few complete power law relations can be found. Many power laws break down when scale is too large or too small (Bak, 1996). In this case, a scaling range always appears in a log-log plot for a scaling relation. And thus, a fractal approach based on Minkowski's dilation curves can be employed to identify the urban boundaries (Montero *et al*, 2021; Tannier *et al*, 2011; Tannier and Thomas, 2013). Third, self-affine relation. Sometimes, a scaling breaking in a power-law distribution implies a self-affine process and pattern and thus takes on a pseudo exponential distribution (Chen and Feng, 2012). In this case, we can use both exponential distribution and scaling breaking point to find characteristic searching radius. These approaches will be explored and the related questions will be discussed in a future research.

## 5 Conclusions

The basic algorithm of defining urban boundary has been developed, but how to find an objective searching radius is a pending problem. Fractal approach proved to be suitable for determining urban envelope for bi-scaling patterns of urban form. In this paper, the characteristic length of spatial search is proposed to define urban boundary. If we can reveal an exponential relationship between searching radius and corresponding measurements, or if we can decompose a power law based on searching radius into two exponential functions, we can find a characteristic length from the scale parameters. The main conclusions can be reached as follows. **First, the scale parameter of the exponential model represents the characteristic searching radius for definition of urban boundary.** Generally speaking, the relation between spatial searching radii and the numbers of land patch clusters follows exponential decay law. The reciprocal of the decay coefficient is regarded as the scale parameter, which is just the characteristic length of spatial search. Using the characteristic radius, we can determine an urban agglomeration or define urban boundary more effectively and objectively. **Second, the average value and double standard deviation can be employed to find average characteristic searching radius for a system of cities.** Different cities at different times have different characteristic searching radius. For individual cities, we can use the characteristic radius of a concrete city. For a system of cities in a region, however, we must find a common characteristic radius by averaging after removing outliers. If a characteristic radius of a cities is less than the average value minus double standard deviation or greater than the average value plus double deviation, it should be deleted as an outlier. Step by step, we can eliminate all the exceptional values and calculate the final average characteristic radius. **Third, other models of spatial search with characteristic length can be transformed into exponential model to give the characteristic searching radius.** Not all the cities follow the exponential decay law in the process of city clustering. If the relation between searching radii and cluster numbers follow the power law with single scaling process, the urban boundary cannot be found objectively. If the relation satisfy a function with characteristic scale, e.g., normal function, logarithmic function, lognormal function, gamma function, and so on, we can convert the function into the form of an exponential function and find the characteristic searching radius. As soon as the characteristic radius is worked out, an urban agglomeration or urban boundary can be determined in a proper way.

### Acknowledgements

This research was sponsored by the National Natural Science Foundation of China (Grant No. 42171192). The support is gratefully acknowledged.

## Supplementary files

**[Supplementary File S1]** *The datasets of searching radii and clusters numbers for 24 Chinese cities in 2000* (Excel). In this file, the datasets of 2000 year and the related calculation results are provided. The final tables are also show in the datasheet.

**[Supplementary File S2]** *The datasets of searching radii and clusters numbers for 24 Chinese cities in 2010* (Excel). In this file, the datasets of 2010 year and the related calculation results are provided. The final tables and statistic experiments are also show in the datasheet.

# Appendix 1—Beijing clusters based on different searching radii (2010)

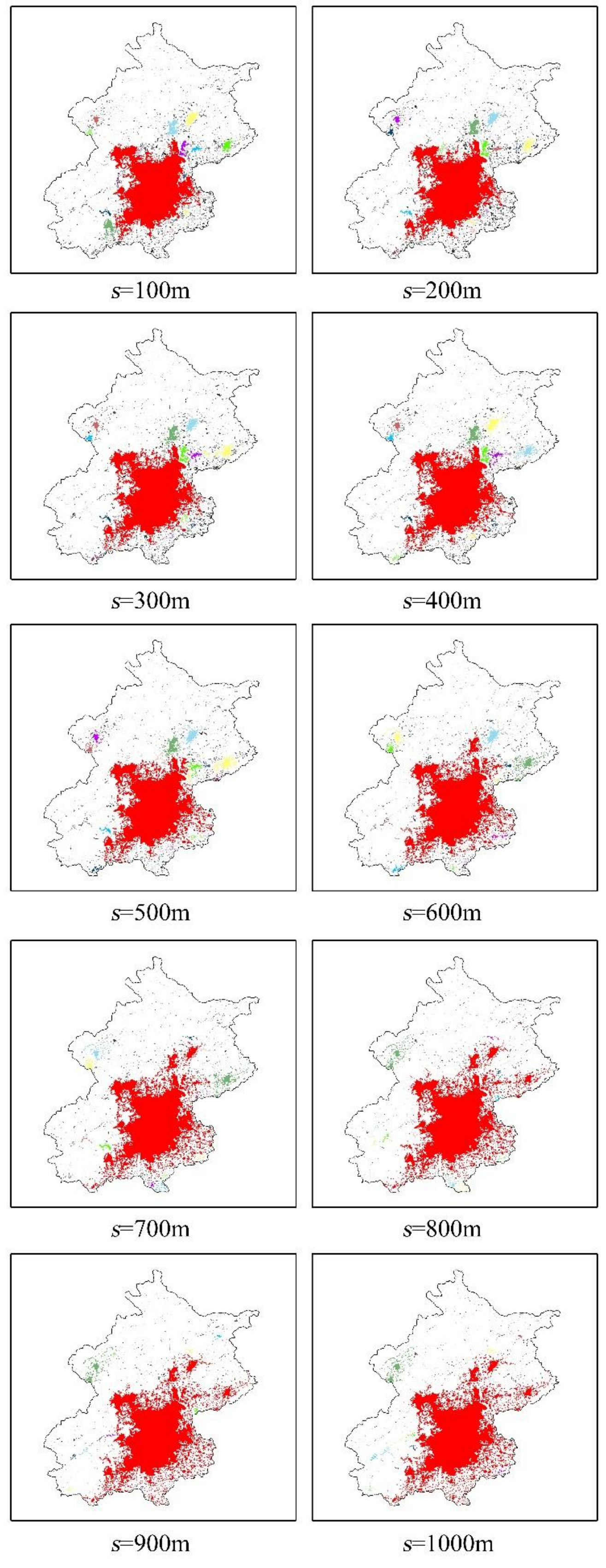

**Figure A1 The clusters of Beijing's urban agglomeration based on different searching radius (separated patterns)**

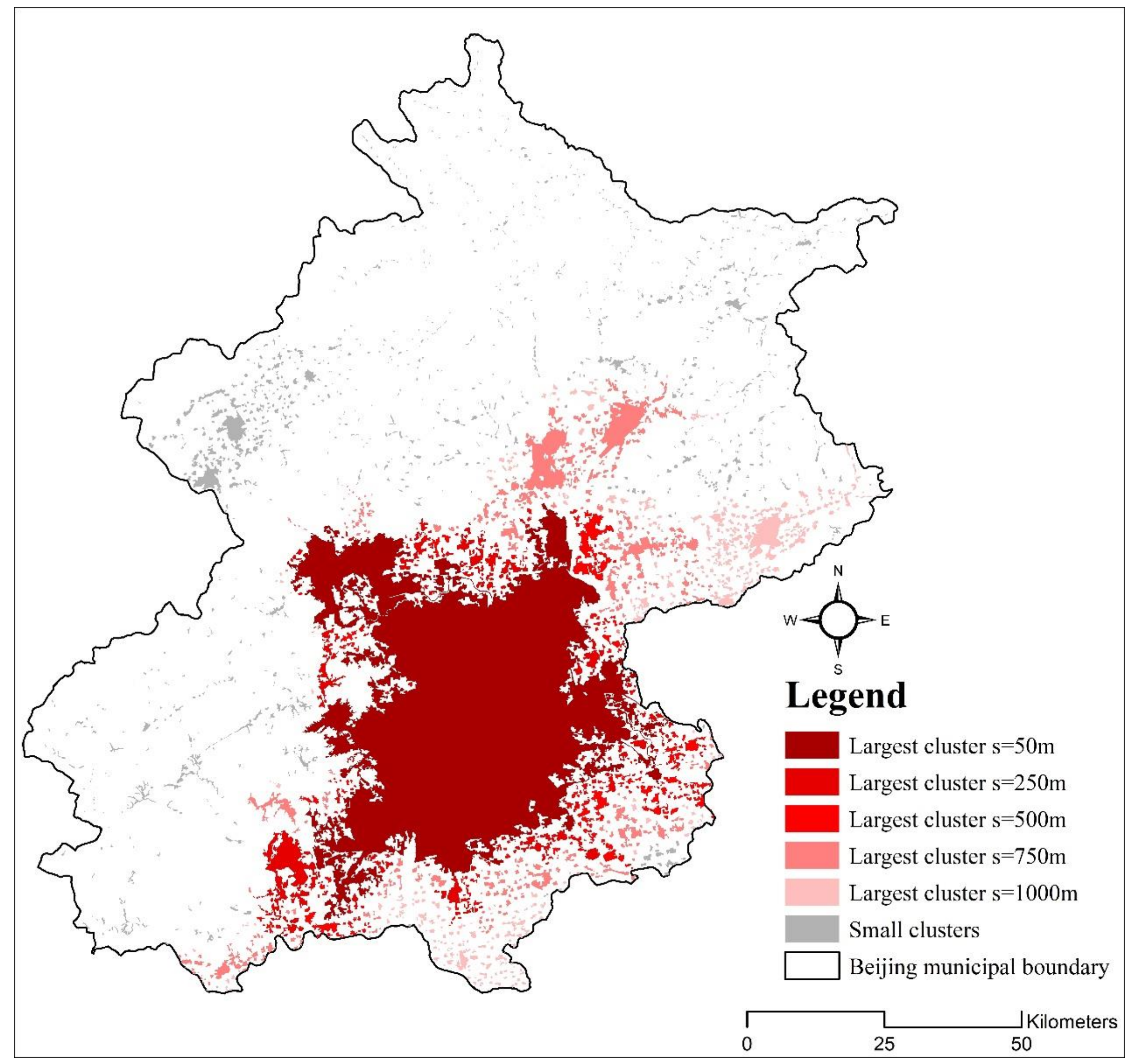

**Figure A2 The clusters of Beijing's urban agglomeration based on different searching radius (Superimposed pattern)**

# Appendix 2—The systems of cities and towns in Jing-Jin-Ji region of China

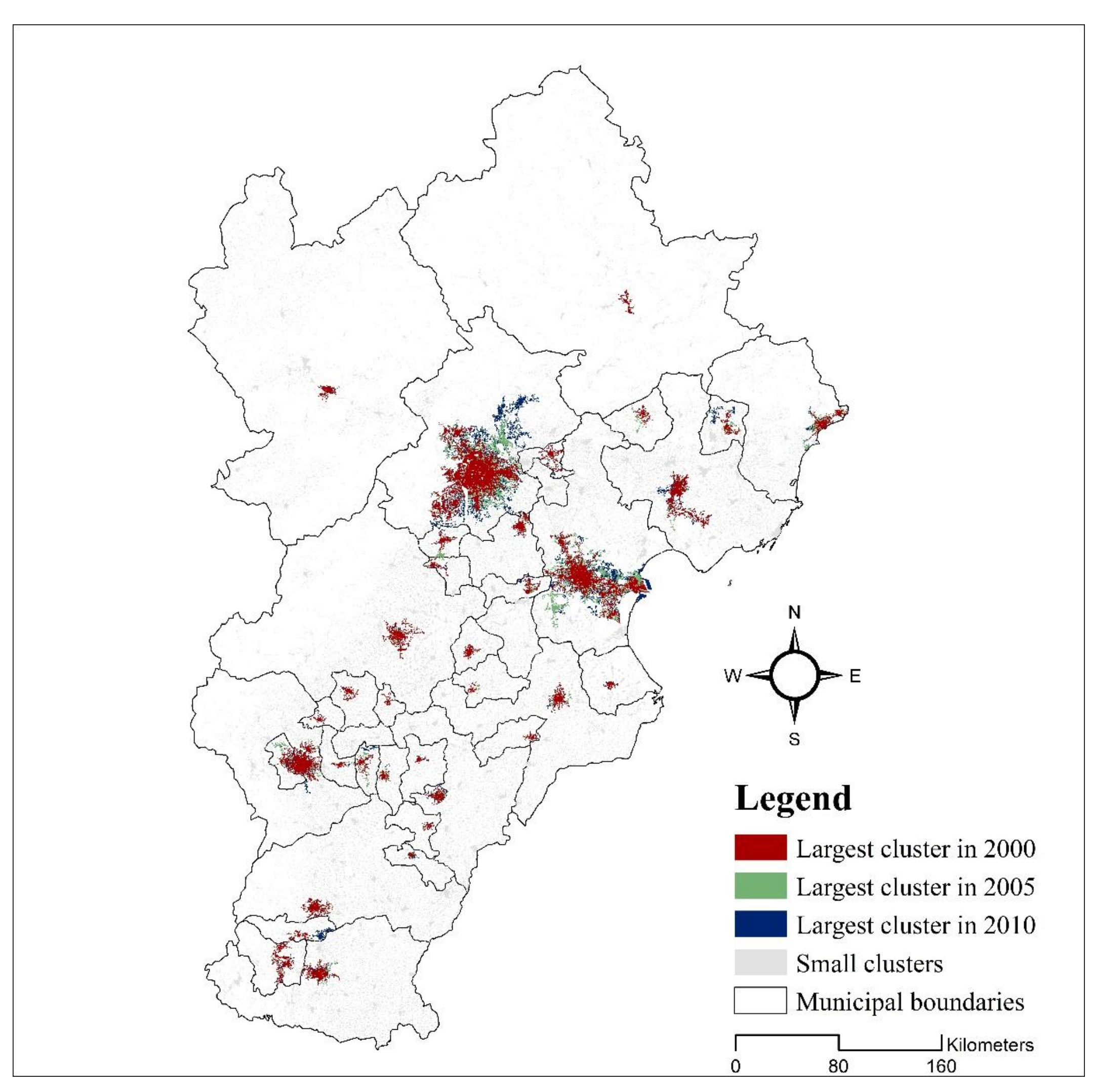

## Appendix 3—The decay coefficients, characteristic searching radii, and goodness of fit of Jing-Jin-Ji system of cities

| City | 2000 | | | 2005 | | | 2010 | | |
|---|---|---|---|---|---|---|---|---|---|
| | $b$ | $s_0$ | $R^2$ | $b$ | $s_0$ | $R^2$ | $b$ | $s_0$ | $R^2$ |
| **Anguo** | 0.002091 | 478.2194 | 0.9815 | 0.002074 | 482.0655 | 0.9826 | 0.002131 | 469.3131 | 0.9828 |
| **Baoding** | 0.001881 | 531.6702 | 0.9984 | 0.001940 | 515.4485 | 0.9984 | 0.001990 | 502.6293 | 0.9985 |
| **Bazhou** | 0.001647 | 607.0195 | 0.9910 | 0.001750 | 571.5378 | 0.9878 | 0.002188 | 456.9559 | 0.9808 |
| **Beijing** | 0.002119 | 471.9511 | 0.9994 | 0.002368 | 422.2940 | 0.9984 | 0.002729 | 366.3731 | 0.9984 |
| **Botou** | 0.001680 | 595.2899 | 0.9967 | 0.001738 | 575.3568 | 0.9971 | 0.001845 | 541.9359 | 0.9974 |
| **Cangzhou** | 0.001772 | 564.4511 | 0.9969 | 0.001801 | 555.3676 | 0.9971 | 0.001917 | 521.7741 | 0.9974 |
| **Chengde** | 0.001124 | 889.6798 | 0.9992 | 0.001165 | 858.2158 | 0.9992 | 0.001181 | 846.5634 | 0.9993 |
| **Dingzhou** | 0.002082 | 480.2106 | 0.9908 | 0.002134 | 468.5543 | 0.9922 | 0.002196 | 455.3302 | 0.9895 |
| **Gaobeidian** | 0.002093 | 477.6993 | 0.9576 | 0.002189 | 456.7534 | 0.9629 | 0.002411 | 414.7853 | 0.9621 |
| **Gaocheng** | 0.002150 | 465.0345 | 0.9934 | 0.002266 | 441.2548 | 0.9943 | 0.002441 | 409.7140 | 0.9928 |
| **Handan** | 0.001846 | 541.8339 | 0.9966 | 0.002006 | 498.4697 | 0.9966 | 0.002189 | 456.7426 | 0.9967 |
| **Hejian** | 0.001709 | 585.2021 | 0.9957 | 0.001767 | 565.9756 | 0.9959 | 0.001749 | 571.7135 | 0.9965 |
| **Hengshui** | 0.001887 | 529.8914 | 0.9957 | 0.001987 | 503.1921 | 0.9956 | 0.002079 | 480.9873 | 0.9977 |
| **Huanghua** | 0.002321 | 430.9048 | 0.9817 | 0.002325 | 430.1719 | 0.9830 | 0.002439 | 410.0397 | 0.9860 |
| **Jinzhou** | 0.002445 | 408.9475 | 0.9961 | 0.003038 | 329.1444 | 0.9981 | 0.003219 | 310.6132 | 0.9972 |
| **Jizhou** | 0.001348 | 741.8193 | 0.9952 | 0.001375 | 727.2594 | 0.9941 | 0.001341 | 745.9429 | 0.9923 |
| **Langfang** | 0.001711 | 584.5939 | 0.9960 | 0.001737 | 575.8528 | 0.9967 | 0.001965 | 509.0170 | 0.9967 |
| **Luquan** | 0.003048 | 328.0825 | 0.9986 | 0.003268 | 305.9994 | 0.9983 | 0.003595 | 278.1609 | 0.9990 |
| **Nangong** | 0.001392 | 718.4533 | 0.9933 | 0.001537 | 650.4204 | 0.9903 | 0.001597 | 626.2028 | 0.9907 |
| **Qian'an** | 0.003069 | 325.8163 | 0.9951 | 0.003589 | 278.5973 | 0.9964 | 0.004004 | 249.7421 | 0.9966 |
| **Qinhuangdao** | 0.001747 | 572.4701 | 0.9992 | 0.001856 | 538.8632 | 0.9985 | 0.002063 | 484.7142 | 0.9984 |
| **Renqiu** | 0.002329 | 429.3267 | 0.9950 | 0.002356 | 424.4385 | 0.9945 | 0.002325 | 430.0616 | 0.9937 |
| **Sanhe** | 0.003514 | 284.5509 | 0.9970 | 0.003654 | 273.6597 | 0.9969 | 0.004335 | 230.6794 | 0.9939 |
| **Shahe** | 0.002423 | 412.7810 | 0.9925 | 0.002597 | 385.0894 | 0.9930 | 0.002764 | 361.8166 | 0.9928 |
| **Shenzhou** | 0.001286 | 777.4866 | 0.9927 | 0.001456 | 686.9067 | 0.9954 | 0.001731 | 577.7408 | 0.9956 |
| **Shijiazhuang** | 0.002038 | 490.7084 | 0.9967 | 0.002115 | 472.7405 | 0.9974 | 0.002341 | 427.2531 | 0.9966 |
| **Tangshan** | 0.002686 | 372.3012 | 0.9990 | 0.002752 | 363.3882 | 0.9990 | 0.002959 | 337.9401 | 0.9987 |
| **Tianjin** | 0.002806 | 356.4120 | 0.9977 | 0.003201 | 312.4214 | 0.9977 | 0.003383 | 295.6242 | 0.9971 |
| **Wuan** | 0.002269 | 440.7886 | 0.9853 | 0.002367 | 422.4660 | 0.9889 | 0.002616 | 382.3133 | 0.9794 |
| **Xingtai** | 0.001518 | 658.9464 | 0.9975 | 0.001612 | 620.3984 | 0.9978 | 0.001732 | 577.4773 | 0.9973 |
| **Xinji** | 0.001684 | 593.9871 | 0.9969 | 0.001709 | 585.1828 | 0.9964 | 0.001933 | 517.2225 | 0.9957 |
| **Xinle** | 0.001739 | 575.0715 | 0.9953 | 0.001792 | 558.0866 | 0.9974 | 0.002047 | 488.5642 | 0.9972 |
| **Zhangjiakou** | 0.000789 | 1266.9277 | 0.9994 | 0.000836 | 1195.6389 | 0.9990 | 0.000872 | 1146.7506 | 0.9990 |
| **Zhuozhou** | 0.002178 | 459.0393 | 0.9908 | 0.002253 | 443.8220 | 0.9944 | 0.002272 | 440.0895 | 0.9928 |
| **Zunhua** | 0.002212 | 452.0322 | 0.9972 | 0.002571 | 388.8955 | 0.9906 | 0.002687 | 372.1908 | 0.9916 |
| **Average** | 0.002018 | 539.9886 | 0.9937 | 0.002148 | 510.9694 | 0.9941 | 0.002322 | 476.9993 | 0.9934 |
| **Stdev** | 0.000567 | 182.6956 | 0.0077 | 0.000645 | 175.5432 | 0.0069 | 0.000737 | 172.8801 | 0.0075 |
| **Average*** | 0.002110 | 490.3683 | 0.9932 | 0.002244 | 471.9630 | 0.9937 | 0.002343 | 443.0722 | 0.9930 |
| **Stdev*** | 0.000413 | 88.3327 | 0.0083 | 0.000583 | 108.1967 | 0.0071 | 0.000488 | 82.5355 | 0.0080 |

**Note**: The notation is as follows: $b$—decay coefficient, $s_0$—characteristic searching radius, $R^2$—goodness of fit. Stdev means “standard deviation”. The average values and standard deviations with asterisk “*” are based on the datasets from which the outliers are removed.